\documentclass[twocolumn,american,english]{revtex4}
\usepackage[T1]{fontenc}
\usepackage[latin9]{inputenc}
\usepackage{color}
\usepackage{array}
\usepackage{amsmath}
\usepackage{esint}

\makeatletter

\providecommand{\tabularnewline}{\\}

\@ifundefined{textcolor}{}
{%
 \definecolor{BLACK}{gray}{0}
 \definecolor{WHITE}{gray}{1}
 \definecolor{RED}{rgb}{1,0,0}
 \definecolor{GREEN}{rgb}{0,1,0}
 \definecolor{BLUE}{rgb}{0,0,1}
 \definecolor{CYAN}{cmyk}{1,0,0,0}
 \definecolor{MAGENTA}{cmyk}{0,1,0,0}
 \definecolor{YELLOW}{cmyk}{0,0,1,0}
}

\makeatother

\usepackage{babel}
\begin{document}

\title{Analytic gradients for natural orbital functional theory}

\author{Ion Mitxelena$^{1,2}$, Mario Piris$^{1,2,3}$ }

\address{$^{1}$Kimika Fakultatea, Euskal Herriko Unibertsitatea (UPV/EHU),
P.K. 1072, 20080 Donostia, Spain.}

\address{$^{2}$Donostia International Physics Center (DIPC), 20018 Donostia,
Spain.}

\address{$^{3}$IKERBASQUE, Basque Foundation for Science, 48013 Bilbao, Spain.\vspace{0.8cm}
}
\begin{abstract}
The analytic energy gradients with respect to nuclear motion are derived
for natural orbital functional (NOF) theory. The resulting equations
do not require to resort to linear-response theory, so the computation
of NOF energy gradients is analogous to gradient calculations at the
Hartree-Fock level of theory. The structures of 15 spin-compensated
systems, composed by first- and second-row atoms, are optimized employing
the conjugate gradient algorithm. As functionals, two orbital-pairing
approaches were used, namely, the fifth and sixth Piris NOFs (PNOF5
and PNOF6). For the latter, the obtained equilibrium geometries are
\textcolor{black}{compared} with coupled cluster singles and doubles
(CCSD) calculations and accurate empirical data.
\end{abstract}
\maketitle

\section{Introduction}

Since in 1958 Bratoz \cite{Bratoz} derived for first time the analytic
gradient for the restricted Hartree-Fock (HF) case, the development
and applications of analytical gradients has been of great interest
for chemistry and physics \cite{Pulay1977}. Energy gradients are
primarily employed to locate and characterize critical points on the
energy surface in electronic structure theory, \textcolor{black}{especially}
minima and saddle points, and calculate rovibrational spectroscopic
constants and energy levels. The direct analytical calculation of
energy derivatives from the wavefunction is computationally more complex
than the numerical calculation, but offers greater speed and accuracy.
In fact, that is why it has been invested much effort in the development
of analytic energy derivatives for many well-known electronic structure
methods, such as configuration interaction (CI) \cite{Handy1984,OsamuraCIanalgrads},
density cumulant functional theory (DCFT) \cite{Sokolovdcft}, Moller-Plesset
perturbation theory (MP2) \cite{PopleMP2analgrads}, or coupled cluster
(CC) theory including different number of excitations, as recently
Gauss and Stanton did for the full singles, doubles and triples (CCSDT)
method \cite{GaussgradsCCSDT}.

From the very beginning \textcolor{black}{there have }been many attempts
to use the Hellmann-Feynman theorem for calculating energy gradients
\cite{Bakken-HF,Huber-HF,Vianna-HF}, since this approach allows to
compute them by using exclusively one-electron operators. It is important
to note that the theorem is only valid if all parameters entering
the involved density matrices are invariant with respect to nuclear
distortion. Unfortunately, this condition is met solely in the complete
basis set limit because the location of atomic orbitals (AO) is not
important. To achieve accurate results, calculations require the contribution
from two-electron terms, which are in turn the bottleneck of the analytic
energy gradient computation. In this work, the method proposed in
Ref. \cite{Dupuis-comp-2} has been followed to compute efficiently
derivatives of the two-electron integrals.

By reconstructing the second-order reduced density matrix (2-RDM)
$\mathbf{D}$ from the knowledge of the first-order reduced density
matrix (1-RDM) $\mathbf{\Gamma}$, in its spectral representation,
the electronic energy functional can be explicitly written in terms
of the natural orbitals (NOs) and corresponding occupation numbers
(ONs), leading to the natural orbital functional (NOF) theory (NOFT)
\cite{Piris2007,Piris2014a,Pernal2016}. In the last decade, Piris
and collaborators have proposed \cite{Piris2006,Piris2013b} a series
of NOFs known in the literature as PNOFi (i=$\overline{1,6}$), which
have been able to reproduce, in many cases, a degree of accuracy comparable
to those provided by high-level standard electronic structure methods
\cite{Ruiperez2013,Ramos-Cordoba2015,Cioslowski2015c,Lopez2015,Piris2016,Mitxelena}.
In the present article, we develop the analytic energy gradients for
the NOFT, and in particular for PNOF. To our knowledge, this is the
first direct analytical calculation of the energy derivatives with
respect to nuclear motion in NOFT. Perhaps the only precedent is the
derivation of analytical gradients in the IBCS theory, which can be
considered as a NOFT \cite{Piris1998}.

No iterative procedure is needed in order to evaluate the derivative
expressions, therefore, the presented here theory is analogous to
the gradient computation at the HF level of theory. Our methodology
allows the calculation of analytic energy gradients corresponding
to a correlated method at low computational cost, in comparison with
standard wavefunction based methods that must resort to linear-response
theory in order to evaluate the energy derivatives with respect to
nuclear distortions.

This paper is organized as follows. The basic equations involving
NOFT are introduced in section \ref{sub:The-NOFT}, followed by the
development of general expressions for the energy gradients with respect
to nuclear motion in section \ref{sub: Analytic-gradients-in-NOFT},
and analytic gradients for PNOF in section \ref{sub:Analytic-gradients-for-PNOF}.
The next section \ref{sec:Computational-aspects} is dedicated to
discussing the computational aspects related to energy gradient calculations.
In section \ref{sec:results}, we compare the optimized structures
of 15 spin-compensated systems at the PNOF5 and PNOF6 levels of theory
with respect to the corresponding coupled cluster singles and doubles
{[}CCSD{]} results, by using the correlation consistent triple-zeta
(cc-pVTZ) basis set developed by Dunning and coworkers \cite{Dunning1989}.
Accurate empirical geometries \cite{bak} are included in order to
carry out a statistical analysis.

\section{Theory}

\subsection{Natural Orbital Functional Theory \label{sub:The-NOFT}}

In the Born-Oppenheimer approximation, the total energy of an \textit{N}-electron
molecule can be cast as the sum of the nuclear and electronic energies,
\begin{equation}
E=E_{nuc}+E_{el}={\displaystyle \sum_{A<B}}\frac{Z_{A}Z_{B}}{R_{AB}}+E_{el},\label{Etotal}
\end{equation}

being the electronic energy ($E_{el}$)\textcolor{black}{, }an exactly
and explicitly known functional of the 1- and 2-RDMs, 
\begin{equation}
E_{el}=\sum\limits _{ik}\Gamma_{ki}\mathcal{H}_{ki}+\sum\limits _{ijkl}D_{klij}\left\langle kl|ij\right\rangle .\label{Eelec_0}
\end{equation}
In Eq. (\ref{Eelec_0}), $\mathcal{H}_{ki}$ are the one-electron
matrix elements of the core-Hamiltonian, whereas $\left\langle kl|ij\right\rangle $
are the two-electron integrals of the Coulomb interaction. In the
following, all representations used are assumed to refer to the basis
in which the one-matrix $\mathbf{\Gamma}$ is diagonal. $E_{el}$
can be then expressed in terms of the NOs and their ONs by means of
a reconstruction functional $\mathbf{D}\left[\mathbf{\Gamma}\right]$
\cite{Piris2007}, which leads to the following general expression
of a NOF 
\begin{equation}
E_{el}=\sum_{i}n_{i}\mathcal{H}_{ii}+V_{ee}\left[N,\left\{ n_{i}\right\} ,\left\{ \phi_{i}\left(\boldsymbol{\mathrm{x}}\right)\right\} \right],\label{Eelec_1}
\end{equation}
where $n_{i}$ stands for the ON of the NO $\phi_{i}\left(\mathbf{x}\right)$,
and $V_{ee}$ represents the electron-electron interaction energy
functional. Here, ${\bf x\equiv}\left({\bf r,s}\right)$ stands for
the combined spatial and spin coordinates, ${\bf r}$ and ${\bf s}$,
respectively. The spin-orbitals $\left\{ \phi_{i}\left({\bf x}\right)\right\} $
constitute a complete orthonormal set of single-particle functions,
\begin{equation}
<\phi_{k}|\phi_{i}>=\int d{\bf x}\phi_{k}^{\ast}\left({\bf x}\right)\phi_{i}\left({\bf x}\right)=\delta_{ki}\label{ortho}
\end{equation}
with an obvious meaning of the Kronecker delta $\delta_{ki}$.

NOFT is an electron correlation method without wavefunction, therefore,
taking into account the exact energy expression (\ref{Eelec_0}),
we must ensure that the RDMs are $N$-representable \cite{Coleman1963,Percus}.
In the case of the 1-RDM, the ensemble $N$-representability conditions
\cite{Coleman1963} reduce to $0\leq n_{i}\leq1$ and $\sum_{i}n_{i}=N$,
often referred to as boundary and normalization conditions of the
occupancies. However, due to the complexity of the necessary and sufficient
conditions for ensuring that the 2-RDM corresponds to an $N$-particle
wavefunction \cite{Mazziotti-Nrep}, we must settle for tractable
necessary conditions for the $N$-representability of the 2-RDM in
any approximation for $V_{ee}$. Note that the one-electron part of
the functional (\ref{Eelec_1}) is exactly defined in terms of the
1-RDM, so the $N$-representability of the energy functional (\ref{Eelec_1})
relies on the corresponding constraints for the 2-RDM \cite{Mazziotti2007}
due to exact expression (\ref{Eelec_0}). 

The procedure for the minimization of the energy (\ref{Eelec_1})
requires optimizing with respect to the ONs and the NOs, separately.
The method of Lagrange multipliers is used to ensure the orthonormality
requirement (\ref{Eelec_1}) for the NOs, and the normalization condition
for the ONs. The latter can additionally be expressed by means of
auxiliary variables in order to automatically enforce the N-representability
bounds of the 1-RDM. Hence, the auxiliary functional $\Lambda\left[N,\left\{ n_{i}\right\} ,\left\{ \phi_{i}\right\} \right]$
is given by
\begin{equation}
\begin{array}{c}
\Lambda=E_{el}-\mu\left({\displaystyle \sum_{i}}n_{i}-N\right)-{\displaystyle \sum_{ki}}\lambda_{ik}\left(\left\langle \phi_{k}|\phi_{i}\right\rangle -\delta_{ki}\right).\end{array}\label{Lagrangian}
\end{equation}

By making (\ref{Lagrangian}) stationary with respect to the NOs and
ONs, we obtain the following system of Eqs.:

\begin{equation}
\frac{\partial E_{el}}{\partial n_{i}}=\mathcal{H}_{ii}+\frac{\partial V_{ee}}{\partial n_{i}}=\mu,\label{equation_for_occupations}
\end{equation}
\begin{equation}
\frac{\partial E_{el}}{\partial\phi_{i}^{*}}=n_{i}\hat{\mathcal{H}}\phi_{i}+\frac{\partial V_{ee}}{\partial\phi_{i}^{*}}=\sum_{k}\lambda_{ki}\phi_{k}.\label{orbital_EULER_equation}
\end{equation}

Eq. (\ref{equation_for_occupations}) is obtained holding the orbitals
fixed, whereas the set of the orbital Euler Eqs. (\ref{orbital_EULER_equation})
is satisfied for a fixed set of occupancies. At present, the procedure
of solving simultaneously Eqs. (\ref{equation_for_occupations}) and
(\ref{orbital_EULER_equation}) is carried out by the iterative diagonalization
method developed by Piris and Ugalde \cite{Piris2009a}, which has
proven to be a powerful tool for attaining the solutions in NOFT.

\subsection{Analytic gradients in NOFT\label{sub: Analytic-gradients-in-NOFT}}

Assume all NOs are real and expand them in a fixed basis set, $\phi_{i}\left(\mathbf{x}\right)=\sum_{\upsilon}\mathcal{C}_{\upsilon i}\zeta_{\upsilon}\left(\mathbf{x}\right)$,
then, the electronic energy (\ref{Eelec_0}) can be rewritten as
\begin{equation}
E_{el}={\displaystyle \sum_{\mu\upsilon}\Gamma_{\mu\upsilon}\mathcal{H}_{\mu\upsilon}+\sum_{\mu\upsilon\eta\delta}D_{\mu\eta\upsilon\delta}\left\langle \mu\eta|\upsilon\delta\right\rangle },\label{Eel_2}
\end{equation}

where $\Gamma_{\mu\upsilon}$ and $D_{\mu\eta\upsilon\delta}$ are
respectively the 1- and 2-RDM given in the atomic orbital (AO) representation,
\begin{equation}
\begin{array}{c}
\Gamma_{\mu\upsilon}={\displaystyle \sum_{i}}n_{i}\mathcal{C}_{\mu i}\mathcal{C}_{\upsilon i},\\
D_{\mu\eta\upsilon\delta}=\sum\limits _{klij}D_{klij}\mathcal{C}_{\mu k}\mathcal{C}_{\eta l}\mathcal{C}_{\upsilon i}\mathcal{C}_{\delta j}.
\end{array}\label{RDM_ao_1}
\end{equation}
Then, the derivative of the total energy (\ref{Etotal}) with respect
to the coordinate $x$ of nucleus $A$ is given by
\begin{equation}
\frac{dE}{dx_{A}}=\frac{\partial E_{el}}{\partial x_{A}}+\frac{\partial E_{nuc}}{\partial x_{A}}+\sum_{\mu j}\frac{\partial E_{el}}{\partial\mathcal{C}_{\mu j}}\frac{\partial\mathcal{C}_{\mu j}}{\partial x_{A}}+\sum_{i}\frac{\partial E_{el}}{\partial n_{i}}\frac{\partial n_{i}}{\partial x_{A}},\label{derivative}
\end{equation}

where $\partial E_{el}/\partial x_{A}$ and $\partial E_{nuc}/\partial x_{A}$
represents the derivative of all terms with explicit dependence on
the nuclear coordinate $x_{A}$, whereas the last two terms in Eq.
(\ref{derivative}) arise from the implicit dependence of the orbital
coefficients and ONs on geometry, respectively.

The electronic energy (\ref{Eel_2}) presents explicit dependence
on the nuclear motion via one- and two-electron integrals, due to
the dependence of the AOs on the geometry, namely,
\begin{equation}
{\displaystyle \frac{\partial E_{el}}{\partial x_{A}}=\sum_{\mu\upsilon}\Gamma_{\mu\upsilon}\frac{\partial\mathcal{H}_{\mu\upsilon}}{\partial x_{A}}+}{\displaystyle \sum_{\mu\upsilon\eta\delta}}D_{\mu\eta\upsilon\delta}\frac{\partial\left\langle \mu\eta|\upsilon\delta\right\rangle }{\partial x_{A}}.\label{explicit-deriv}
\end{equation}

The first term in Eq. (\ref{explicit-deriv}) is the negative \textit{Hellmann-Feynman
force} \cite{H.Hellmann,R.P.Feynman}. The second term, which contains
the derivatives of the two-electron integrals, is the bottleneck for
calculating the analytical gradient.

Regarding the contribution from the NO coefficients, combining Eq.
(\ref{orbital_EULER_equation}) with the chain rule, is not difficult
to obtain the next formula:
\begin{equation}
\frac{\partial E_{elec}}{\partial\mathcal{C}_{\mu j}}=2\sum_{\upsilon i}\mathcal{S}_{\mu\upsilon}\mathcal{C}_{\upsilon i}\lambda_{ij},\label{derivative_coefficients}
\end{equation}
where $\mathcal{S_{\mu\upsilon}}$ is the overlap matrix $\left\langle \mu|\upsilon\right\rangle $.
At the same time, the response of NO coefficients to nuclear motion
can be computed from the orthonormality relation (\ref{ortho}) in
the AO representation ($\mathbf{C^{\dagger}SC}=\mathbf{1}$) \cite{Pulay},
indeed,

\begin{equation}
2\:{\displaystyle \sum_{\mu\upsilon}}\frac{\partial\mathcal{C}_{\mu j}}{\partial x_{A}}\mathcal{S_{\mu\upsilon}}\mathcal{C}_{\upsilon i}=-{\displaystyle \sum_{\mu\upsilon}}\mathcal{C}_{\mu j}\frac{\partial\mathcal{S_{\mu\upsilon}}}{\partial x_{A}}\mathcal{C}_{\upsilon i}.\label{derivative_pulay}
\end{equation}

Combining then Eqs. (\ref{derivative_coefficients}) and (\ref{derivative_pulay}),
and taking into account the contribution from different indexes, we
obtain the total contribution from the NO coefficients to the gradient,
which is known as the \textit{density force}:
\begin{equation}
\sum_{\mu j}\frac{\partial E_{el}}{\partial\mathcal{C}_{\mu j}}\frac{\partial\mathcal{C}_{\mu j}}{\partial x_{A}}=-{\displaystyle \sum_{\mu\upsilon}\lambda_{\mu\upsilon}}\frac{\partial\mathcal{S_{\mu\upsilon}}}{\partial x_{A}},\label{derivative_coefficients_total}
\end{equation}
where
\begin{equation}
{\displaystyle \lambda_{\mu\upsilon}=\sum_{ij}\mathcal{C}_{\mu j}}\lambda_{ji}\mathcal{C}_{\upsilon i}.
\end{equation}
The last term of Eq. (\ref{derivative}) does not bring any contribution
to the gradient, since deriving the normalization condition ($\sum_{i}n_{i}=N$)
of the ONs, one obtains
\begin{equation}
\sum_{i}\frac{\partial n_{i}}{\partial x_{A}}=0.\label{derivative_occupations}
\end{equation}
Hence, combining (\ref{derivative_occupations}) and (\ref{equation_for_occupations}),
brings about a contribution to the gradient equal to zero:
\begin{equation}
\sum_{i}\frac{\partial E_{el}}{\partial n_{i}}\frac{\partial n_{i}}{\partial x_{A}}=\mu\sum_{i}\frac{\partial n_{i}}{\partial x_{A}}=0.\label{derivative_occupations_2}
\end{equation}
Finally, bringing together Eqs. (\ref{explicit-deriv}) and (\ref{derivative_coefficients_total})
with the nuclear contribution $\partial E_{nuc}/\partial x_{A}$,
we obtain the expression for NOF analytic gradients, namely,

\begin{equation}
\begin{array}{c}
{\displaystyle \frac{dE}{dx_{A}}}={\displaystyle \sum_{\mu\upsilon}}\Gamma_{\mu\upsilon}\dfrac{\partial\mathcal{H}_{\mu\upsilon}}{\partial x_{A}}+{\displaystyle \sum_{\mu\upsilon\eta\delta}}D_{\mu\eta\upsilon\delta}\dfrac{\partial\left\langle \mu\eta|\upsilon\delta\right\rangle }{\partial x_{A}}\\
\qquad+{\displaystyle \frac{\partial E_{nuc}}{\partial x_{A}}}-{\displaystyle \sum_{\mu\upsilon}}\lambda_{\mu\upsilon}\dfrac{\partial\mathcal{S_{\mu\upsilon}}}{\partial x_{A}}.
\end{array}\label{NOF-analy-grads}
\end{equation}
The spin orbitals are direct products $\left\vert \phi_{i}\right\rangle =$
$\left\vert \varphi_{p}\right\rangle \otimes$ $\left\vert \sigma\right\rangle $,
so $\left\{ \phi_{i}\left(\mathbf{x}\right)\right\} $ may be split
into two subsets: $\left\{ \varphi_{p}^{\alpha}\left(\mathbf{r}\right)\alpha\left(\mathbf{s}\right)\right\} $
and $\left\{ \varphi_{p}^{\beta}\left(\mathbf{r}\right)\beta\left(\mathbf{s}\right)\right\} $.
Given a set of $2R$ spin-orbitals $\left\{ \phi_{i}|i=1,...,2R\right\} $,
we have two sets of $R$ orthonormal spatial functions, $\left\{ \varphi_{p}^{\alpha}\left(\mathbf{r}\right)\right\} $
and $\left\{ \varphi_{p}^{\beta}\left(\mathbf{r}\right)\right\} $,
such that in general the first set is not orthogonal to the second
one. Nevertheless, the original set
\begin{align*}
\phi_{2p-1}\left(\mathbf{x}\right) & =\varphi_{p}^{\alpha}\left(\mathbf{r}\right)\alpha\left(\mathbf{s}\right),\text{ \ \ \ \ }p=1,...,R\\
\phi_{2p}\left(\mathbf{x}\right) & =\varphi_{p}^{\beta}\left(\mathbf{r}\right)\beta\left(\mathbf{s}\right),\text{ \ \ \ \ }p=1,...,R
\end{align*}
 continues being orthonormal via the orthogonality of the spin functions
\begin{equation}
\int d\mathbf{s}\alpha^{\ast}\left(\mathbf{s}\right)\beta\left(\mathbf{s}\right)=\int d\mathbf{s}\beta^{\ast}\left(\mathbf{s}\right)\alpha\left(\mathbf{s}\right)=0.
\end{equation}

Since we deal herein only with singlet states, the spin restricted
formulation is employed, in which a single set of orbitals is used
for $\alpha$ and $\beta$ spins: $\varphi_{p}^{\alpha}\left(\mathbf{r}\right)=\varphi_{p}^{\beta}\left(\mathbf{r}\right)=\varphi_{p}\left(\mathbf{r}\right)$.
Similarly as we did above for the spin NOs, let us expand the spatial
NOs $\left\{ \varphi_{p}\right\} $ as a linear combination of atomic
orbitals: $\varphi_{p}\left(\mathbf{r}\right)=\sum_{\upsilon}\mathcal{C}_{\upsilon p}\chi_{\upsilon}\left(\mathbf{r}\right)$.
Consequently, in Eq. (\ref{NOF-analy-grads}), entering magnitudes
become
\begin{equation}
\begin{array}{c}
\Gamma_{\mu\upsilon}=2{\displaystyle \sum_{p}}n_{p}\mathcal{C}_{\mu p}\mathcal{C}_{\upsilon p}\\
D_{\mu\eta\upsilon\delta}=2\sum\limits _{pqrt}(D_{pqrt}^{\alpha\alpha}+D_{pqrt}^{\alpha\beta})\mathcal{C}_{\mu p}\mathcal{C}_{\eta q}\mathcal{C}_{\upsilon r}\mathcal{C}_{\delta t}\\
{\displaystyle \lambda_{\mu\upsilon}=2\sum_{pq}}\mathcal{C}_{\mu q}\lambda_{qp}\mathcal{C}_{\upsilon p}.
\end{array}\label{RDM_ao_2}
\end{equation}

\subsection{Analytic gradients for PNOF\label{sub:Analytic-gradients-for-PNOF}}

In this work, we use a particular reconstruction of the 2-RDM in terms
of the 1-RDM proposed by Piris \cite{Piris2006}. Thus, the electronic
energy for a system with an even number $N$ of electrons is given
by the $\mathcal{JKL}$-only NOF 
\begin{equation}
\begin{array}{c}
E_{el}=2{\displaystyle \sum_{p}n_{p}\mathcal{H}_{pp}}+{\displaystyle \sum_{pq}}\Pi_{qp}\mathcal{L}_{pq}\qquad\qquad\\
{\displaystyle \qquad+\sum_{pq}\left(n_{q}n_{p}-\Delta_{qp}\right)\left(2\mathcal{J}_{pq}-\mathcal{K}_{pq}\right)}.
\end{array}\label{Eelec_2}
\end{equation}

In Eq. (\ref{Eelec_2}), $\mathcal{J}_{pq}=\left\langle pq|pq\right\rangle $
and $\mathcal{K}_{pq}=\left\langle pq|qp\right\rangle $ are the usual
direct and exchange integrals, respectively, whereas $\mathcal{L}_{pq}=\left\langle pp|qq\right\rangle $
is the exchange and time-inversion integral \cite{Piris1999}. $\Delta$
and $\Pi$ are the auxiliary matrices introduced in reference \cite{Piris2006},
which exclusively depend on the ONs. The so called (2,2)-positivity
conditions \cite{Mazziotti-Nrep} for the $N$-representability of
the 2-RDM provide bounds for the off-diagonal terms of matrices $\Delta$
and $\Pi$ \cite{Piris2010a}, whereas the conservation of the total
spin allows to determine the diagonal elements \textcolor{black}{as}
$\Delta_{pp}=n_{p}^{2}$ and $\Pi_{pp}=n_{p}$ \cite{Piris2009}.

For real orbitals, $\mathcal{L}_{pq}$ reduces to $\mathcal{K}_{pq}$
, so the energy functional (\ref{Eelec_2}) becomes a $\mathcal{JK}$-only
NOF,
\begin{equation}
\begin{array}{c}
E_{el}=2{\displaystyle \sum_{p}n_{p}\mathcal{H}_{pp}}+{\displaystyle 2\sum_{pq}\left(n_{q}n_{p}-\Delta_{qp}\right)\mathcal{J}_{pq}}\\
{\displaystyle -\sum_{pq}\left(n_{q}n_{p}-\Delta_{qp}-\Pi_{qp}\right)\mathcal{K}_{pq}}.
\end{array}\label{Eelec_3}
\end{equation}
Accordingly, the analytical gradients for PNOF are given by Eq. (\ref{NOF-analy-grads})
together with the 1- and 2-RDM defined in (\ref{RDM_ao_2}), where
the latter is now expressed as
\begin{equation}
\begin{array}{c}
D_{\mu\eta\upsilon\delta}=\sum\limits _{pq}\left[2\left(n_{q}n_{p}-\Delta_{qp}\right)\mathcal{C}_{\mu p}\mathcal{C}_{\upsilon p}\mathcal{C}_{\eta q}\mathcal{C}_{\delta q}\right.\\
\qquad\qquad\left.-\left(n_{q}n_{p}-\Delta_{qp}-\Pi_{qp}\right)\mathcal{C}_{\mu p}\mathcal{C}_{\delta p}\mathcal{C}_{\eta q}\mathcal{C}_{\upsilon q}\right].
\end{array}\label{2RDM_ao_PNOF}
\end{equation}
Note that the four-index summation appearing in Eq. (\ref{RDM_ao_2})
for the 2-RDM is reduced to only two in Eq. (\ref{2RDM_ao_PNOF}),
due to the two-index nature of the PNOF reconstruction that leads
to a $\mathcal{JKL}$-only NOF \cite{Piris2006}.

\subsubsection*{Orbital pairing approaches\label{sub:Orbital-pairing-approaches}}

Recently, the electron-pairing approach has been exploited in PNOF
theory. Two approximations with pairing restrictions have been proposed
so far, namely, the independent-pair model (PNOF5) \cite{Piris2011}
and an interacting-pair approximation (PNOF6) \cite{Piris2014}. Extended
versions of both approaches have been proposed \cite{Piris2013e,Piris2015a},
too. For them, the orbital space $\Omega$ is partitioned into $F=N/2$
subspaces $\{\Omega{}_{g}\}$. The subspaces are considered mutually
disjoint $\left(\Omega_{g1}\cap\Omega_{g2}=\textrm{Ø}\right)$, i.e.,
each orbital belongs only to one subspace $\Omega_{g}$. Each subspace
contains one orbital $g$ below the Fermi level ($F$), and $N_{g}$
coupled orbitals above it, which is reflected in additional sum rules
for the ONs in each $\Omega{}_{g}$, namely,
\begin{equation}
\sum_{p\in\Omega_{g}}n_{p}=1.\label{sumrule_n}
\end{equation}

Note that $N_{g}=1$ corresponds to the simplest formulations, PNOF5
or PNOF6, whereas $N_{g}>1$ leads to different extended formulations
of both. Taking into account the spin, each subspace contains solely
an electron pair, and the normalization condition ($2\sum_{p}n_{p}=N$)
is automatically fulfilled. Moreover, we look for orbitals belonging
to each subspace $\Omega_{g}$, which yield the minimum energy for
the functional of Eq. (\ref{Eelec_3}). However, the actual orbitals
that satisfy the pairing conditions (\ref{sumrule_n}) are not constrained
to remain fixed along the orbital optimization process. Consequently,
the pairing scheme of the orbitals is allowed to vary along the optimization
process till the most favorable orbital interactions are found.

In accordance to these new constraints, we may associate new Lagrange
multipliers $\{\mu_{g}\}$ with the $F$ pairing conditions (\ref{sumrule_n}),
instead of the chemical potential $\mu$. It has been suggested \cite{Piris2015}
that the smallest $\mu_{g}$ can be then identified as the chemical
potential of an open system. The auxiliary functional $\Lambda$ (\ref{Lagrangian})
may be in turn redefined by the formula
\begin{equation}
\Lambda=E_{el}-{\displaystyle 2\sum_{g=1}^{F}}\mu_{g}\sum_{p\in\Omega_{g}}(n_{p}-1)-{\displaystyle 2\sum_{qp}}\lambda_{pq}\left(\left\langle \varphi_{q}|\varphi_{p}\right\rangle -\delta_{qp}\right).\label{auxiliari_lagrangian_pairing}
\end{equation}

The partial derivative $\left(\partial E/\partial n_{p}\right)$,
holding the spatial NOs $\{\varphi_{g}\}$ fixed, are now given by
the expressions
\begin{equation}
\frac{\partial E_{el}}{\partial n_{p}}=2\mathcal{H}_{pp}+\frac{\partial V_{ee}}{\partial n_{p}}=2\mu_{g},\:\forall p\in\Omega_{g}.\label{equation_for_occupations_pairing}
\end{equation}
Regarding the analytical gradient equation for orbital pairing approaches,
the Eq. (\ref{derivative_occupations}) fulfills independently for
each orbital subspace $g$ due to relation (\ref{sumrule_n}), 
\begin{equation}
\sum_{p\in\Omega_{g}}\frac{\partial n_{p}}{\partial x_{A}}=0,\label{derivative_occupations_g}
\end{equation}

thereby the contribution to the gradient becomes zero for each subspace
\begin{equation}
\sum_{g}\sum_{p\in\Omega_{g}}\frac{\partial E_{el}}{\partial n_{p}}\frac{\partial n_{p}}{\partial x_{A}}=2\sum_{g}\mu_{g}\sum_{p\in\Omega_{g}}\frac{\partial n_{p}}{\partial x_{A}}=0.\label{derivative_occupations_3}
\end{equation}
In consequence, the analytical energy gradients (\ref{NOF-analy-grads})
remain unmodified for orbital pairing approaches.

\section{Computational aspects\label{sec:Computational-aspects}}

The Eq. (\ref{NOF-analy-grads}) implies that we do not require an
iterative procedure for evaluating the derivative of the total energy
with respect to the coordinate $x_{A}$. The gradient can be efficiently
computed by first calculating the quantities $\Gamma{}_{\mu\upsilon}$,
$D_{\mu\eta\upsilon\delta}$, and $\lambda_{\mu\upsilon}$, subsequently
contracting by derivatives of the integrals.

In contrast to what happens in other post-HF theories, our methodology
allows the calculation of analytic energy gradients by the simple
evaluation without resorting to the linear-response theory. Our gradient
computation is therefore analogous to that which is performed at the
HF level of theory with the corresponding savings of computational
time. Indeed, the PNOF analytic gradient reduces to the HF expression
after removing $\Delta$ and $\Pi$ matrices in Eq. (\ref{2RDM_ao_PNOF}),
i.e., the two-electron cumulant matrix \cite{Piris2006}. Consequently,
as it happens in the HF case, the bottleneck of gradient evaluation
is the computation of the two-electron contribution, since 12 gradient
components arise from each two-electron integral \cite{Dupuis-comp-2}.\textcolor{black}{{}
In this sense, our approach is similar to the projected Hartree-Fock
method that recovers a significant portion of static correlation too
\cite{Schutski2014}.}

Overall, the calculation scales nominally as $M^{5}$ ($M$ being
the number of basis set functions) due to the $pq$-linkage in the
auxiliary matrices of the PNOF $D_{\mu\eta\upsilon\delta}$, given
by Eq. (\ref{2RDM_ao_PNOF}). However, in case of pairing approximations,
the auxiliary matrices could contain a lot of zeros corresponding
to neglecting ONs of the higher NOs in energy. For instance, in case
of simplest pairing, PNOF5 or PNOF6, the number of involved NOs with
non-zero occupancies is equal to the the number of electrons $N$,
therefore, the summations by $p$ and $q$, in Eq.(\ref{2RDM_ao_PNOF}),
are up to $N$ instead of $M$, and the scaling reduces from $M^{5}$
to $N\cdot M^{4}$. Obviously, factorized PNOF auxiliary matrices
$\Delta$ and $\Pi$, i.e., $\Delta_{qp}=\Delta_{q}\Delta_{p}$ and
$\Pi_{qp}=\Pi_{q}\Pi_{p}$, could reduce the scaling to $M^{4}$.
In this case, we could make the summations by $p$ and $q$ before
contracting by derivatives of the integrals, in a similar way to what
one does in the HF approximation.

In practice, the scaling is also reduced by applying a previous screening
of two-electron integrals based on Schwarz' inequality \cite{horn},
\textcolor{black}{especially in the case of large systems where the
smallness of most two-electron integrals allows to skip their evaluation.
In any case, the basis set employed determines the computational}
time instead of the number of geometrical degrees of freedom.

In the present implementation, as there is no constrain regarding
the nuclear coordinates of the system, we use the well-known nonlinear
conjugate gradient (CG) method \cite{fletcher} to locate ground state
equilibrium geometries. This algorithm associates conjugacy properties
with the steepest descent method, so that both efficiency and reliability
are achieved, as reflected in the results reported in the next section.
The main advantage is that the method requires only gradient evaluations
and does not use much storage, because the search direction is acquired
from linear combinations of the gradient obtained in the previous
iteration. Its main drawback is that the search direction is not necessarily
down. Herein, the studied systems are simple molecules with starting
configurations close to the optimized geometries, therefore we have
no doubt that they are equilibrium geometries. \textcolor{black}{For
diatomic molecules herein studied, the harmonic frequency analyses
have already been done in previous works \cite{Piris2014,Piris2013e,Piris2013c}.
Nevertheless, to be sure of having reached a minimum in the other
systems, it is required to compute the Hessian (matrix of second derivatives)
in addition to the gradient. Note that it is possible to avoid the
problems inherent to the analytic calculation of the Hessian, such
as storage issues, solving coupled perturbed equations, or computing
the large amount of two-electron integral second derivatives, by a
numerical differentiation of analytic gradients \cite{Bykov2015}.}

\section{results\label{sec:results}}

In this section, we carry out a NOF study of the ground-state equilibrium
geometries for a selected set of spin-compensated molecules. This
set includes the following 15 systems: HF, H$_{2}$O, NH$_{3}$, CH$_{4}$,
N$_{2}$, CO, HOF, HNO, H$_{2}$CO, HNNH, H$_{2}$CCH$_{2}$, HCCH,
HCN, HNC, and O$_{3}$. As functionals, two orbital-pairing approaches
were used, namely, PNOF5 and PNOF6. Both functionals, including their
extended versions, take into account most of the non-dynamical effects,
but also the important part of dynamical electron correlation corresponding
to the intrapair interactions \cite{Piris2011,Piris2014,Piris2016,Ramos-Cordoba2015,Ruiperez2013,Cioslowski2015c,Lopez2015}.
PNOF5 does not describe correlation between electron pairs at all,
while PNOF6 includes mostly non-dynamic interpair correlation. 

\textcolor{black}{We use HF geometries as starting points to PNOF
optimizations. For comparison, we have included high-quality empirical
equilibrium structures obtained from least-squares fits involving
experimental rotational constants and theoretical vibrational corrections
\cite{bak}. Furthermore, the corresponding CCSD \cite{GaussgradsCCSDT}
values are included. All calculations are carried out using the correlation-consistent
polarized triple-zeta (cc-pVTZ) basis set developed by Dunning and
coworkers \cite{Dunning1989}, which are suitable in correlated calculations
\cite{bak}.}

\begin{table*}
\caption{Errors in the equilibrium bonds (in pm) at PNOF5, PNOF6, and CCSD
levels of theory calculated by using the cc-pVTZ basis set with respect
to empirical structural data. $\overline{\Delta}$ and $\overline{\Delta}_{abs}$
correspond to the mean signed error and mean absolute error, respectively.
\bigskip{}
}

\noindent \begin{centering}
\begin{tabular}{>{\raggedright}m{2.5cm}>{\raggedright}p{2.5cm}c>{\centering}p{2cm}>{\centering}p{2cm}>{\centering}p{3cm}}
Molecule & Bond & PNOF5 & PNOF6 & CCSD\cite{GaussgradsCCSDT} & $\quad\;$EMP.\cite{bak}\tabularnewline
\hline 
HF & H\textemdash{}F  & -0.2  & -0.3 & -0.3 & \ 91.7\tabularnewline
H$_{2}$O  & O\textemdash{}H & \ 0.1  & -0.5 & -0.2 & \ 95.8\tabularnewline
NH$_{3}$ & N\textemdash{}H & \ 0.6 & -0.3 & -0.3 & 101.2\tabularnewline
CH$_{4}$ & C\textemdash{}H  & \ 1.5 & -0.5 & -0.1 & 108.6\tabularnewline
N$_{2}$ & N\textemdash{}N & -0.7 & -1.4 & -0.4 & 109.8\tabularnewline
CO & C\textemdash{}O & -1.1 & -1.5 & -0.3 & 112.8\tabularnewline
HNO & N\textemdash{}O & \ 0.0 & -1.3 & -0.9 & 120.9\tabularnewline
 & H\textemdash{}N & -0.7 & -2.1 & -0.3 & 105.2\tabularnewline
H$_{2}$CO  & C\textemdash{}O & \ 0.2 & -1.1 & -0.5 & 120.5\tabularnewline
 & C\textemdash{}H & \ 0.4 & -1.1 & -0.4 & 110.1\tabularnewline
HNNH & N\textemdash{}N & -0.1 & -1.2 & -0.7 & 124.6\tabularnewline
 & N\textemdash{}H & \ 0.1 & -1.6 & -0.4 & 102.9\tabularnewline
H$_{2}$CCH$_{2}$ & C\textemdash{}C & \ 0.9 & -0.3 & -0.4 & 133.1\tabularnewline
 & C\textemdash{}H & \ 1.1 & -0.7 & -0.4 & 108.1\tabularnewline
HCCH & C\textemdash{}C & -0.1 & -1.0 & -0.4 & 120.4\tabularnewline
 & C\textemdash{}H & \ 0.7 & -0.7 & -0.4 & 106.1\tabularnewline
HCN & C\textemdash{}N & -0.5 & -1.3 & -0.4 & 115.3\tabularnewline
 & C\textemdash{}H & \ 0.5 & -0.8 & -0.6 & 106.5\tabularnewline
HNC & C\textemdash{}N & -2.3 & -1.3 & -0.4 & 116.9\tabularnewline
 & N\textemdash{}H & -1.3 & -1.0 & -0.4 & \ 99.5\tabularnewline
HOF & O\textemdash{}F & \ 3.6 & \ 2.4 & -1.9 & 143.4\tabularnewline
 & H\textemdash{}O & -0.3 & -1.9 & -0.5 & \ 96.8\tabularnewline
O$_{3}$ & O\textemdash{}O & \ 2.6 & -3.5 & -3.6 & \ 127.2{*}\tabularnewline
\hline 
$\overline{\Delta}$ &  & \ 0.2 & -1.0 & -0.6 & \tabularnewline
$\overline{\Delta}_{abs}$ &  & \ 0.8 & \ 1.2 & \ 0.6 & \tabularnewline
\end{tabular}\label{table-bonds}
\par\end{centering}

\noindent \raggedright{}{*}Geometry extracted from Ref. \cite{nist}.
\end{table*}

\begin{table*}
\caption{Errors in the equilibrium bond angles (in degs) at PNOF5, PNOF6, and
CCSD levels of theory calculated by using the cc-pVTZ basis set with
respect to empirical structural data. $\overline{\Delta}$ and $\overline{\Delta}_{abs}$
correspond to the mean signed error and mean absolute error, respectively.
\bigskip{}
}

\noindent \begin{centering}
\begin{tabular}{>{\raggedright}p{2.5cm}>{\raggedright}p{2.5cm}c>{\centering}p{2cm}>{\centering}p{2cm}>{\centering}p{3cm}}
\hline 
Molecule & Bond angle & PNOF5 & PNOF6 & CCSD\cite{GaussgradsCCSDT} & EMP.\cite{bak}\tabularnewline
\hline 
H$_{2}$O  & H\textemdash{}O\textemdash{}H & \ 0.23 & \ 0.04 & -0.47 & 104.51\tabularnewline
NH$_{3}$ & H\textemdash{}N\textemdash{}H & \ 0.45 & -0.89 & -0.89 & 107.25\tabularnewline
HOF & H\textemdash{}O\textemdash{}F & -0.27 & -0.22 & \ 0.43 & \ 97.94\tabularnewline
HNO & H\textemdash{}N\textemdash{}O & -0.53 & \ 0.21 & \ 0.00 & 108.27\tabularnewline
H$_{2}$CO  & H\textemdash{}C\textemdash{}O & -0.09 & \ 0.07 & \ 0.29 & 121.63\tabularnewline
HNNH & H\textemdash{}N\textemdash{}N & \ 0.82 & \ 1.07 & -0.04 & 106.36\tabularnewline
H$_{2}$CCH$_{2}$ & H\textemdash{}C\textemdash{}C & -0.15 & \ 0.00 & \ 0.03 & 121.43\tabularnewline
O$_{3}$ & O\textemdash{}O\textemdash{}O & -3.44 & \ 0.09 & \ 1.57 & \ 116.70{*}\tabularnewline
\hline 
$\overline{\Delta}$ &  & -0.37 & \ 0.05 & \ 0.12 & \tabularnewline
$\overline{\Delta}_{abs}$ &  & \ 0.75 & \ 0.33 & \ 0.47 & \tabularnewline
\end{tabular}\label{table-angles}
\par\end{centering}

\noindent \raggedright{}{*}Geometry extracted from Ref. \cite{nist}.
\end{table*}

Tables \ref{table-bonds} and \ref{table-angles} show respectively
the errors in bond lengths and bond angles obtained for the selected
set of molecules at PNOF5, PNOF6 and CCSD levels of theory, along
with the empirical equilibrium structures. Note that reported NOF
results involve the simplest coupling ($N_{g}=1$) in our calculations,
so each orbital below the Fermi level is coupled with a single orbital
above it.

A survey of both tables \ref{table-bonds} and \ref{table-angles}
reveals that both NOFs employed here, PNOF5 and PNOF6, provide ground-state
\textcolor{black}{equilibrium structures comparable to those of the
CCSD. For PNOF5, the corresponding mean absolute errors $\overline{\Delta}_{abs}$
are 0.75 degs and 0.8 pm for bond angles and bond lengths, respectively,
which are slightly above }0.47 degs and 0.6 pm obtained by using the
CCSD method. PNOF6 performs relatively worse for bond distances ($\overline{\Delta}_{abs}=1.2$
pm), but \textcolor{black}{it provides the best }bond angles ($\overline{\Delta}_{abs}=0.33$
degs), even better than the behavior of CCSD. The slight differences
with respect to CCSD are mainly due to the HNC, HOF and O$_{3}$ molecules,
for which the largest errors are observed.

\textcolor{black}{It is worth noting that the systems studied in the
current work can be well described by independent-pair approximations
since they do not present delocalized electrons. For the latter, it
is well-known that approaches like PNOF5 predict symmetry-breaking
artifacts \cite{Piris2014,Ramos-Cordoba2015}. On the other hand,
the $\mathcal{JKL}$-only functional PNOF6 includes interactions between
the electron pairs, but to the detriment of the correlation energy
that recovers, which is smaller than that obtained with PNOF5 in the
presented herein systems \cite{Piris2011,Piris2014}. That is why
calculated bond distances decreases when going from PNOF5 to PNOF6,
as happens when going to a lower-energy correlation method in wavefunction-based
theories.}

We note that the independent-pair approximation (PNOF5) underestimates
some inter-atomic distances, while overestimates in other cases, with
a slight tendency to the latter as evidenced by the mean signed value
$\overline{\Delta}$ given in Table \ref{table-bonds}. It is worth
to note that this trend has been observed when perturbative triples
are included in CC theory \cite{bak} for the used basis set. On the
other hand, the inclusion of the interactions between electron pairs
by PNOF6, underestimates the bond distances in all studied cases,
as CCSD consistently does, with the exception of the O\textemdash{}F
length in the HOF molecule. In the case of bond angles, PNOF5 behaves
similarly, but here the trend is slightly to underestimate, whereas
PNOF6 reports practically equal values to experimental data ($\overline{\Delta}=0.05$
degs), according to the results reported in Table \ref{table-angles}.
Obviously, more sample molecules are needed in order to come to a
conclusion.

The case of ozone is remarkable, since none of the methods used in
this work give a satisfactory result for the O\textemdash{}O bond
length in comparison with the experimental value. Although, we should
note that PNOF6 corrects the O\textemdash{}O\textemdash{}O bond angle
obtained by using PNOF5, so the interactions between electron pairs
seem to play an important role in O$_{3}$. Interestingly, for the
employed cc-pVTZ basis set, CCSD(T) is able to correct the CCSD value
and yield a bond distance of 127.6 pm \cite{GaussgradsCCSDT} with
an error of 0.4 pm, despite \textcolor{black}{O$_{3}$}\textcolor{red}{{}
}\textcolor{black}{being }a typical two-configuration system.

\begin{table*}
\caption{\textcolor{black}{Errors in the equilibrium bonds (in pm) at PNOF5,
PNOF5(5), PNOF6, and PNOF6(5) levels of theory calculated by using
the cc-pVTZ basis set with respect to empirical structural data. $\overline{\Delta}$
and $\overline{\Delta}_{abs}$ correspond to the mean signed error
and mean absolute error, respectively.} \bigskip{}
}

\noindent \begin{centering}
\textcolor{black}{}%
\begin{tabular}{>{\raggedright}m{2.5cm}>{\raggedright}p{2.5cm}c>{\centering}p{2cm}>{\centering}p{2cm}>{\centering}p{2cm}}
\textcolor{black}{Molecule} & \textcolor{black}{Bond} & \textcolor{black}{PNOF5} & \textcolor{black}{PNOF5(5)} & \textcolor{black}{PNOF6} & \textcolor{black}{PNOF6(5)}\tabularnewline
\hline 
\textcolor{black}{HF} & \textcolor{black}{H\textemdash{}F } & \textcolor{black}{-0.2 } & \textcolor{black}{-0.5} & \textcolor{black}{-0.3} & \textcolor{black}{-1.2}\tabularnewline
\textcolor{black}{H$_{2}$O } & \textcolor{black}{O\textemdash{}H} & \textcolor{black}{\ 0.1 } & \textcolor{black}{-0.3} & \textcolor{black}{-0.5} & \textcolor{black}{-0.9}\tabularnewline
\textcolor{black}{NH$_{3}$} & \textcolor{black}{N\textemdash{}H} & \textcolor{black}{\ 0.6} & \textcolor{black}{\ 0.1} & \textcolor{black}{-0.3} & \textcolor{black}{-1.3}\tabularnewline
\textcolor{black}{CH$_{4}$} & \textcolor{black}{C\textemdash{}H } & \textcolor{black}{\ 1.5} & \textcolor{black}{-0.4} & \textcolor{black}{-0.5} & \textcolor{black}{-0.2}\tabularnewline
\textcolor{black}{N$_{2}$} & \textcolor{black}{N\textemdash{}N} & \textcolor{black}{-0.7} & \textcolor{black}{-0.9} & \textcolor{black}{-1.4} & \textcolor{black}{-2.2}\tabularnewline
\textcolor{black}{CO} & \textcolor{black}{C\textemdash{}O} & \textcolor{black}{-1.1} & \textcolor{black}{-1.4} & \textcolor{black}{-1.5} & \textcolor{black}{-1.8}\tabularnewline
\textcolor{black}{HNO} & \textcolor{black}{N\textemdash{}O} & \textcolor{black}{\ 0.0} & \textcolor{black}{-0.5} & \textcolor{black}{-1.3} & \textcolor{black}{-1.8}\tabularnewline
 & \textcolor{black}{H\textemdash{}N} & \textcolor{black}{-0.7} & \textcolor{black}{-1.4} & \textcolor{black}{-2.1} & \textcolor{black}{-2.1}\tabularnewline
\textcolor{black}{H$_{2}$CO } & \textcolor{black}{C\textemdash{}O} & \textcolor{black}{\ 0.2} & \textcolor{black}{-0.1} & \textcolor{black}{-1.1} & \textcolor{black}{-1.4}\tabularnewline
 & \textcolor{black}{C\textemdash{}H} & \textcolor{black}{\ 0.4} & \textcolor{black}{-0.3} & \textcolor{black}{-1.1} & \textcolor{black}{-0.8}\tabularnewline
\textcolor{black}{HNNH} & \textcolor{black}{N\textemdash{}N} & \textcolor{black}{-0.1} & \textcolor{black}{-0.4} & \textcolor{black}{-1.2} & \textcolor{black}{-1.7}\tabularnewline
 & \textcolor{black}{N\textemdash{}H} & \textcolor{black}{\ 0.1} & \textcolor{black}{\ 0.5} & \textcolor{black}{-1.6} & \textcolor{black}{-1.4}\tabularnewline
\textcolor{black}{H$_{2}$CCH$_{2}$} & \textcolor{black}{C\textemdash{}C} & \textcolor{black}{\ 0.9} & \textcolor{black}{\ 0.6} & \textcolor{black}{-0.3} & \textcolor{black}{-0.8}\tabularnewline
 & \textcolor{black}{C\textemdash{}H} & \textcolor{black}{\ 1.1} & \textcolor{black}{\ 0.4} & \textcolor{black}{-0.7} & \textcolor{black}{-0.5}\tabularnewline
\textcolor{black}{HCCH} & \textcolor{black}{C\textemdash{}C} & \textcolor{black}{-0.1} & \textcolor{black}{-0.8} & \textcolor{black}{-1.0} & \textcolor{black}{-1.7}\tabularnewline
 & \textcolor{black}{C\textemdash{}H} & \textcolor{black}{\ 0.7} & \textcolor{black}{\ 0.1} & \textcolor{black}{-0.7} & \textcolor{black}{-0.6}\tabularnewline
\textcolor{black}{HCN} & \textcolor{black}{C\textemdash{}N} & \textcolor{black}{-0.5} & \textcolor{black}{-0.8} & \textcolor{black}{-1.3} & \textcolor{black}{-2.0}\tabularnewline
 & \textcolor{black}{C\textemdash{}H} & \textcolor{black}{\ 0.5} & \textcolor{black}{-0.1} & \textcolor{black}{-0.8} & \textcolor{black}{-0.7}\tabularnewline
\textcolor{black}{HNC} & \textcolor{black}{C\textemdash{}N} & \textcolor{black}{-2.3} & \textcolor{black}{-1.3} & \textcolor{black}{-1.3} & \textcolor{black}{-1.7}\tabularnewline
 & \textcolor{black}{N\textemdash{}H} & \textcolor{black}{-1.3} & \textcolor{black}{-0.6} & \textcolor{black}{-1.0} & \textcolor{black}{-0.9}\tabularnewline
\textcolor{black}{HOF} & \textcolor{black}{O\textemdash{}F} & \textcolor{black}{\ 3.6} & \textcolor{black}{\ 3.3} & \textcolor{black}{\ 2.4} & \textcolor{black}{-0.1}\tabularnewline
 & \textcolor{black}{H\textemdash{}O} & \textcolor{black}{-0.3} & \textcolor{black}{-0.9} & \textcolor{black}{-1.9} & \textcolor{black}{-1.9}\tabularnewline
\textcolor{black}{O$_{3}$} & \textcolor{black}{O\textemdash{}O} & \textcolor{black}{\ 2.6} & \textcolor{black}{\ 1.4} & \textcolor{black}{-3.5} & \textcolor{black}{-3.8{*}}\tabularnewline
\hline 
\textcolor{black}{$\overline{\Delta}$} &  & \textcolor{black}{\ 0.2} & \textcolor{black}{-0.2} & \textcolor{black}{-1.0} & \textcolor{black}{-1.4}\tabularnewline
\textcolor{black}{$\overline{\Delta}_{abs}$} &  & \textcolor{black}{\ 0.8} & \textcolor{black}{\ 0.7} & \textcolor{black}{\ 1.2} & \textcolor{black}{\ 1.4}\tabularnewline
\end{tabular}\textcolor{black}{\label{table-bonds-1}}
\par\end{centering}

\noindent \raggedright{}\textcolor{black}{{*} For this molecule 3
orbitals are considered in each subspace.}
\end{table*}

\begin{table*}
\caption{\textcolor{black}{Errors in the equilibrium bond angles (in degs)
at PNOF5, PNOF5(5), PNOF6, and PNOF6(5) levels of theory calculated
by using the cc-pVTZ basis set with respect to empirical structural
data. $\overline{\Delta}$ and $\overline{\Delta}_{abs}$ correspond
to the mean signed error and mean absolute error, respectively.} \bigskip{}
}

\noindent \begin{centering}
\textcolor{black}{}%
\begin{tabular}{>{\raggedright}p{2.5cm}>{\raggedright}p{2.5cm}cc>{\centering}p{2cm}>{\centering}p{2cm}}
\hline 
\textcolor{black}{Molecule} & \textcolor{black}{Bond angle} & \textcolor{black}{PNOF5} & \textcolor{black}{PNOF5(5)} & \textcolor{black}{PNOF6} & \textcolor{black}{PNOF6(5)}\tabularnewline
\hline 
\textcolor{black}{H$_{2}$O } & \textcolor{black}{H\textemdash{}O\textemdash{}H} & \textcolor{black}{\ 0.23} & \textcolor{black}{\ 0.43} & \textcolor{black}{\ 0.04} & \textcolor{black}{\ 0.76}\tabularnewline
\textcolor{black}{NH$_{3}$} & \textcolor{black}{H\textemdash{}N\textemdash{}H} & \textcolor{black}{\ 0.45} & \textcolor{black}{-0.54} & \textcolor{black}{-0.89} & \textcolor{black}{\ 0.46}\tabularnewline
\textcolor{black}{HOF} & \textcolor{black}{H\textemdash{}O\textemdash{}F} & \textcolor{black}{-0.27} & \textcolor{black}{-0.25} & \textcolor{black}{-0.22} & \textcolor{black}{\ 0.58}\tabularnewline
\textcolor{black}{HNO} & \textcolor{black}{H\textemdash{}N\textemdash{}O} & \textcolor{black}{-0.53} & \textcolor{black}{\ 0.07} & \textcolor{black}{\ 0.21} & \textcolor{black}{\ 0.31}\tabularnewline
\textcolor{black}{H$_{2}$CO } & \textcolor{black}{H\textemdash{}C\textemdash{}O} & \textcolor{black}{-0.09} & \textcolor{black}{\ 0.21} & \textcolor{black}{\ 0.07} & \textcolor{black}{\ 0.15}\tabularnewline
\textcolor{black}{HNNH} & \textcolor{black}{H\textemdash{}N\textemdash{}N} & \textcolor{black}{\ 0.82} & \textcolor{black}{\ 1.01} & \textcolor{black}{\ 1.07} & \textcolor{black}{\ 1.01}\tabularnewline
\textcolor{black}{H$_{2}$CCH$_{2}$} & \textcolor{black}{H\textemdash{}C\textemdash{}C} & \textcolor{black}{-0.15} & \textcolor{black}{\ 0.02} & \textcolor{black}{\ 0.00} & \textcolor{black}{\ 0.21}\tabularnewline
\textcolor{black}{O$_{3}$} & \textcolor{black}{O\textemdash{}O\textemdash{}O} & \textcolor{black}{-3.44} & \textcolor{black}{-2.99} & \textcolor{black}{\ 0.09} & \textcolor{black}{0.38{*}}\tabularnewline
\hline 
\textcolor{black}{$\overline{\Delta}$} &  & \textcolor{black}{-0.37} & \textcolor{black}{-0.26} & \textcolor{black}{\ 0.05} & \textcolor{black}{\ 0.48}\tabularnewline
\textcolor{black}{$\overline{\Delta}_{abs}$} &  & \textcolor{black}{\ 0.75} & \textcolor{black}{\ 0.69} & \textcolor{black}{\ 0.33} & \textcolor{black}{\ 0.48}\tabularnewline
\end{tabular}\textcolor{black}{\label{table-angles-1}}
\par\end{centering}

\noindent \raggedright{}\textcolor{black}{{*} For this molecule 3
orbitals are considered in each subspace.}
\end{table*}

\textcolor{black}{One of the possible ways to improve the results
obtained herein is the inclusion of more orbitals in the description
of the electron pairs. For simplicity, consider each orbital $g$
is coupled to a fixed number of orbitals ($N_{g}=N_{c}$), which gives
rise to the functionals PNOF5($N_{c}$) or PNOF6($N_{c}$) as appropriate.
Taking into account that molecules studied here only comprise atoms
of the first and second rows of the periodic table, the inclusion
of 5 more orbitals in each subspace {[}PNOF5(5){]} is suitable to
improve our results. Our results for bond lengths and angles are reported
in Tables \ref{table-bonds-1} and \ref{table-angles-1}, respectively.}

\textcolor{black}{By inspection of Table \ref{table-bonds-1} one
concludes that better description of the intrapair electron correlation
shortens calculated bond lengths. Accordingly, the performance of
PNOF5 improves when the extended approach is employed, whereas in
the case of PNOF6, which tends to underestimate bond distances, calculated
geometries are slightly worse if more orbitals are included in the
description of electron pairs. Table \ref{table-angles-1} shows that
the mean absolute errors differ approximately in 0.1 degs for bond
angles, so the use of extended versions of both functionals does not
affect systematically our results.}

\textcolor{black}{Let us highlight some molecules for which the better
description of the intrapair correlation yields better geometrical
parameters. }In the case of methane, the C\textemdash{}H distance
shortens from 110.1 pm to 108.2 pm, which closely compares to the
experimental value of 108.6 pm. Similarly, the error in HNC bond lengths
reduces from 2.3 pm and 1.3 pm to 1.3 pm and 0.6 pm, respectively,
for the C\textemdash{}N and N\textemdash{}H bonds. It is worth noting
that the only case for which PNOF6 overestimates a bond distance,
the O\textemdash{}F bond in HOF molecule, is corrected using PNOF6(5),
namely, this bond distance shortens from \textcolor{black}{145.9 }pm
to \textcolor{black}{143.4} pm, in outstanding agreement with the
empirical value reported in Table \ref{table-bonds}.

\section{Conclusion}

\textcolor{black}{For first time, we have developed the direct analytical
calculation of the energy derivatives with respect to nuclear motion
in NOFT. Since the energy gradients give much information on potential
energy surfaces and other properties, the study carried out in this
work significantly extends the usefulness of NOFT, which have recently
shown high accuracy calculating dissociation energies \cite{Lopez2015,Piris2016},
electrostatic moments \cite{Mitxelena}, ionization potentials \cite{Piris2012},
etc.}

\textcolor{black}{It is well known that analytical gradients allow
to speed up calculations and avoid numerical errors. The equations
obtained herein allow computing analytic gradients of a correlated
method without solving coupled equations as is the case in most post-HF
methods, for example, in coupled cluster theories, so there is no
need for iterative process to calculate the energy gradient in NOFT.}

\textcolor{black}{By using the nonlinear conjugate gradient method,
we have optimized the structures of 15 spin-compensated molecules
at the PNOF5 and PNOF6 levels of theory, employing the cc-pVTZ basis
set of Dunning. In comparison with the CCSD method, the mean absolute
error in bond distances obtained with PNOF5 differs only in 0.2 pm,
although the difference increases to 0.6 pm when PNOF6 is employed.
Bond angles calculated by using PNOF6 are the most accurate with mean
signed error and mean absolute error equal to 0.05 and 0.33 degs,
respectively. The present article proves the ability of both PNOF5
and PNOF6 to yield geometrical structures at lower computational cost
than other post-HF methods.}

\textcolor{black}{The present study demonstrates the efficiency of
computing energy gradients in NOFT, therefore its calculation in periodic
solids is now affordable. The extension of NOFT to periodic systems
has been done in the past \cite{Piris2005,Sharma2013,Shinohara2015a},
so we expect to achieve a computational efficiency close to that obtained
in HF methodologies \cite{Izmaylov2007}.}

Finally, a comment about the optimization algorithm is mandatory.
The nonlinear conjugate gradient method is often used to solve unconstrained
optimization problems such as the energy minimization studied in this
article. Its main advantage is that it requires only gradient evaluations
and does not use much storage. Its main drawback is that the search
direction is not necessarily down. To be sure of having reached a
minimum or a transition state, or to improve our implementation with
a Newton-like algorithm, we require computing the Hessian (matrix
of second derivatives) in addition to the gradient. A work in this
direction is underway. 

\selectlanguage{american}%

\section*{Acknowledgements}

Financial support comes from Eusko Jaurlaritza (Ref. IT588-13) and
Ministerio de Economía y Competitividad (Ref. CTQ2015-67608-P). The
authors thank for technical and human support provided by IZO-SGI
SGIker of UPV/EHU and European funding (ERDF and ESF). One of us (I.M.)
is grateful to Vice-Rectory for research of the UPV/EHU for the PhD.
grant.

\selectlanguage{english}%
\smallskip{}

\end{document}